\documentstyle[manuscript,eqsecnum,aps,version2]{revtex}
\begin{document}
\title
\bf Aging Random Walks 
\endtitle
\author{Stefan Boettcher$^{1,2}$}
\instit
$^1$Center for Nonlinear Studies, Los Alamos National Laboratory,
Los Alamos, NM 87545\\
$^2$Center for Theoretical Studies of Physical Systems, Clark Atlanta
University, Atlanta, GA 30314\\
\endinstit
\medskip
\centerline{\today}
\abstract
Aging refers to the property of two-time correlation functions to decay
very slowly on (at least) two time scales. This phenomenon has gained
recent attention due to experimental observations of the history
dependent relaxation behavior in amorphous materials (``Glasses'') which
pose a challenge to theorist. Aging signals the breaking of
time-translational invariance and the violation of the fluctuation
dissipation theorem during the relaxation process. But while the origin
of aging in disordered media is profound, and the discussion is clad in
the language of a well-developed theory, systems as simple as a random
walk near a wall can exhibit aging. Such a simple walk serves well to
illustrate the phenomenon and some of the physics behind it.
\medskip

\noindent
PACS number(s): 01.55.+b, 05.40.+j, 02.50.Ey. 
\endabstract

\section{Introduction}

Aging is an important phenomenon observed experimentally in glassy
materials where relaxation behavior depends on the previous history of
the system \cite{AHN}. As a model for a glass Edwards and Anderson
\cite{E+A} introduced an Ising system where the uniform 
coupling $J>0$ between neighboring spins is
replace by random numbers $J_{i,j}$, drawn from some distribution, which
are placed on each bond. The properties of such a disordered system are
quite different when the couplings are drawn from, say, a gaussian
distribution with zero mean, compared to those of the ferro-magnetic
Ising model with uniform couplings $J>0$: While (for $d\ge 2$) the uniform 
Ising model reaches an ordered phase below a critical temperature $T_c$
with a nonzero spontaneous magnetization as order parameter, no such 
order emerges in
the random bond model. Yet, below a certain temperature $T_g$, the ``glass
transition,'' more and more spins loose their mobility and freeze
into place, but without any collectively preferred direction. Thus, while no
macroscopic order parameter emerges, microscopically the state of the
system is ultimately highly (auto-)correlated. The relaxation process
towards such a state is naturally extremely slow due to the inherent
frustration created by the bond distribution, and each spin has to
``negotiate'' its orientation with ever distant neighbors to further
lower the collective energy within their domain.

Consider such a spin glass quenched below the glass transition $T_g$ at 
time (``age'') $t_a=0$ in the presence of a magnetic field. Throughout the 
sample, domains of various pure states develop that grow with characteristic 
size $L(t_a)$ \cite{F+H}. After waiting a time $t_a=t_w$, the magnetic 
field is turned off and the system's response in the form of its 
remnant magnetization is measured. Initially, the response
function is only sensitive to the pure, quasi-equilibrium states in
their distinct domains. But after an additional time scale related to
the waiting time, $t_w$, the response spans entire domains and slows
down when it experiences the off-equilibrium state the sample
possesses as a whole.  In either regime, though, the remnant
magnetization decays very slowly, similar to the sequence of graphs
(see Ref.~\cite{AHN}) in Fig.~2 below.  It has been suggested
\cite{scaling_glasses} that a simple scaling form might hold for
glasses; e.g. that the autocorrelation function for the remnant
magnetization at the age $t_a=t+t_w$ of the experiment is given by
\begin{equation}
C(t+t_w,t_w) = t^{-\mu}f(t/t_w) \quad ,
\label{scaling}
\end{equation}
where the scaling function $f$ is constant for small argument, and 
falls either like a power law (indication of a single aging time scale) 
or like a stretched exponential (indication of multiple aging time scales
\cite{PSAA}) 
for large argument. In any case, time-translational invariance in this 
two-time correlation function $C$ is broken due to the fact that the
relaxation process becomes dependent on its previous history in form of
the waiting time $t_w$.

Similar quantities have been observed in a variety of phenomenological
\cite{Bouch} and theoretical \cite{S+H,Yos,Bar,BoPa3} model systems as 
well, and a debate is raging on the ingredients that are required for a 
model to capture the salient features of glassy materials \cite{BM}. 
While (as we will see) it is easy to produce
aging behavior, the more intricate results from temperature cycling
experiments \cite{slowdyn} are much harder to reproduce, and appear to be 
a much more stringent indicator of the complicated phase-space structure 
(dubbed ``rugged landscapes'' \cite{cnls-proc}) real glasses possess.

In fact, a simple random walk near a (reflecting or absorbing) wall is
well suited to describe the domain-growth picture used above to describe
the observed aging behavior. Consider a walker starting near the wall $n=0$
at $t_a=0$. After a waiting time $t_a=t_w$ she has explored a domain of size
$L(t_w)\sim t_w^{1/2}$ off the wall. The walk now represents the
correlation function for a spin at site $n$
in the above spin glass model, and the wall is the edge of the
(albeit one-sided) domain. In place of magnetization, we measure the probability
$P(n,t+t_w|n,t_w)$ for the walker to return to the same site $n$ she
found herself on at time~$t_w$. Without the wall, $P$ is of course
invariant to shifts in space and time, i. e. $t_w$ is irrelevant and no
aging behavior can be expected. In the presense of the wall, the walker
will venture from the site $n$ for small times $t$ after $t_w$ and find
herself unconstraint (the ``quasi-equilibrium'' state), but when $t\sim t_w$
she is likely to encounter the wall and carry a memory of that
constraint back to the site $n$ for $t\gg t_w$. In the next section, we
will solve this model for a suitably defined two-time correlation
function which indeed shows the expected aging behavior.

Of course, in this model it is the wall that explicitly breaks the
symmetries of the system instead of the dynamics of the
process, and the observed aging behavior appears trivial. On the other hand,
the dynamics of the stochastic annihilation process $A+A\to 0$, which
is similar to a walk at a wall, yields identical results to those
reported below \cite{FKR}, and the distinction between explicit and dynamical
symmetry breaking is not so clear anymore. Finally, the case of an
absorbing wall is particularly revealing and illustrates the
the effect of a process that is slowly dying out (i. e. its norm
decays), while true aging
should only be associated with intrinsic properties of a process that
will sustain itself. This point has lead to some confusion recently.

\section{The Random Walk Model}

In this section, we will calculate the conditional probability
$P(n,t|n_0,t_0)$ for a walker to reach a site $n$ at time $t$, given
that she was at site $n_0$ at some previous time $t_0<t$ in the presense
of either an absorbing or a reflecting wall. Then, we will compute a
simple two-time correlation function (see e. g. Ref.~\cite{S+H} for a
similar definition)
\begin{eqnarray}
C(t+t_w,t_w)=\sum_n P(n,t+t_w|n,t_w) P(n,t_w|0,0)
\label{twotime}
\end{eqnarray}
for a walker to return to a site at time $t_a=t+t_w$, given that she was
at the same site at time $t_w$ after the start of the walk at time
$t_a=0$ near the wall $n=0$. For both boundary conditions, we find that
the walk ages, i. e. shows a scaling behavior according to
Eq.~(\ref{scaling}). Note that the breaking of spatial invariance in $P$
due to
the wall is  crucial for the breaking of time-translational invariance
in $C$: For an unconstraint walk $P$ would be invariant in space and
time, and we would find $P(n,t+t_w|n,t_w)=P(0,t|0,0)$, and with $\sum_n
P(n,t_w|0,0)\equiv 1$, it is $C(t+t_w,t_w)=C(t,0)=P(0,t|0,0)$,
independent of $t_w$. In the presense of the wall, spatial invariance is
broken while time invariance for $P$ still holds, and $C$ merely
simplifies to $C(t+t_w,t_w)=\sum_n P(n,t|n,0) P(n,t_w|0,0)$ which
remains a function of $t_w$.

To simplify the algebra, we consider instead of the walk equation for
$P$ the potential problem
\begin{eqnarray}
\partial_t \phi(r,t)&=&\partial_r^2 \phi(r,t),\quad (r>0,t>0),\cr
\noalign{\medskip}
\phi(r,0)&=&\delta (r-r_0),\cr
\noalign{\medskip}
\phi(0,t)=0\quad &{\rm or}&\quad \partial_r \phi(0,t)=0,
\label{phieq}
\end{eqnarray}
where the two boundary conditions in the last line correspond to an absorbing 
and a reflecting wall, respectively \cite{rem}. We solve for $\phi$, and
identify $P(n,t|r_0,0)=\phi_{r_0}(r,t)$ and
\begin{eqnarray}
C(t+t_w,t_w)=\int_0^{\infty} dr~\phi_r(r,t)~\phi_0(r,t_w).
\label{newc}
\end{eqnarray}
The Eq.~(\ref{phieq}) is easy to solve by converting to an ordinary
differential equation (ODE) in $r$ using a Laplace transform in $t$:
${\tilde \phi}(r,s)=\int_0^{\infty} dt\,e^{-st}\,\phi(r,s)$. 
The ODE has simple exponential solutions in two regions, $0<r<r_0$ and
$r>r_0$, whose four unknown constants are determined by the two boundary
conditions at $r=0$ and $r=\infty$, and by the two matching conditions
at $r=r_0$ where $\phi$ is merely continuous. We find
\begin{eqnarray}
{\tilde\phi}_{r_0}(r,s)={1\over \sqrt{s}}\left[e^{-\sqrt{s}|r-r_0|}\pm
e^{-\sqrt{s}(r+r_0)}\right],
\end{eqnarray}
which is easily inverted using standard tables for Laplace transforms
\cite{A+S}
\begin{eqnarray}
\phi_{r_0}(r,t)={1\over \sqrt{\pi t}}\left[e^{-{(r-r_0)^2\over 4t}}\pm
e^{-{(r+r_0)^2\over 4t}}\right].
\label{result}
\end{eqnarray}
In each case, the upper sign refers to reflecting boundary
conditions, and the lower sign refers to the absorbing case.

\subsection{Reflecting Boundary Conditions}

Here we insert the appropriate forms of $\phi$ in Eq.~(\ref{result}),
using the upper sign case, into
Eq.~(\ref{newc}) and choose $r_0=0$ for convenience \cite{rem2}:
\begin{eqnarray}
C(t+t_w,t_w)&=&{2\over\pi\sqrt{t\,t_w}}\int_0^{\infty}dr \left(1+e^{-{r^2\over
t}}\right) e^{-{r^2\over 4t_w}}\cr
\noalign{\medskip}
&=&{2\over \sqrt{\pi}} \,t^{-{1\over 2}} \, f\left({t\over t_w}\right)\quad
{\rm with}\quad f(x)=1+\sqrt{{x\over 4+x}}.
\end{eqnarray}
Thus, while the two-time correlation function does show the aging
behavior according to Eq.~(\ref{scaling}), its scaling function is
particularly trivial with $f(x\ll 1)\sim 1$ and $f(x\gg 1)\sim 2$, see
Fig.~1.

\subsection{Absorbing Boundary Conditions}

Again, we insert the appropriate forms of $\phi$ in Eq.~(\ref{result}),
using the lower sign case, into Eq.~(\ref{newc}). But with absorbing
boundary conditions, putting the starting position at $r_0=0$ would be
instantly fatal for the walker. (In simulations we use $n=1$ as starting
point.) Instead, we can choose $r_0$ arbitrarily
small and expand to leading order. Since the starting point $r_0$
is irrelevant for the asymptotic behavior (at large times) considered
here \cite{rem2}, we can be sure that higher-order corrections in $r_0$ 
will have to be subdominant:
\begin{eqnarray}
C(t+t_w,t_w)&=&{1\over\pi\sqrt{t\,t_w}}\int_0^{\infty}dr
\left[1-e^{-{r^2\over t}}\right] \left[e^{-{(r-r_0)^2\over 4t_w}}-
e^{-{(r+r_0)^2\over 4t_w}}\right]\cr
\noalign{\medskip}
&\approx&{r_0\over\pi\sqrt{t\,t_w^3}}\int_0^{\infty}dr
\left[1-e^{-{r^2\over t}}\right] re^{-{r^2\over 4t_w}}\cr
\noalign{\medskip}
&=&{r_0\over \sqrt{t_w}}{2\over \pi} \,t^{-{1\over 2}} \,
f\left({t\over t_w}\right)\quad
{\rm with}\quad f(x)={1\over 1+{x\over 4 }}.
\label{cabso}
\end{eqnarray}
In this case, we find more interesting scaling behavior with
$f(x\ll 1)\sim 1$ and $f(x\gg 1)\sim 4/x$. 

But this observed
aging behavior does not consider the effect that a walk actually
disappears when reaching the wall which diminishes the norm of the
distribution ($\phi$ or $P$). Rather, to obtain the intrinsic properties
of an infinite walk near an absorbing wall, we have to properly normalize 
the correlation function. To that end, we consider the two-time correlation 
function $C(t+t_w,t_w|\theta)$ for a walk that reaches the wall (and
disappears) exactly at time $t_a=\theta$, and its generic relation to the
intrinsic two-time correlation function $C^{\rm intr}(t+t_w,t_w)$:
\begin{eqnarray}
C(t+t_w,t_w|\theta)=\cases{0 &$(\theta<t_w+ t)$,\cr
\noalign{\medskip}
C^{\rm intr}(t+t_w,t_w)  &$(\theta\geq t_w+ t)$.}
\end{eqnarray}
These quantities are related to the two-time correlation function given 
in Eq.~(\ref{cabso}) (which is usually the one that is simulated by
averaging over walks of any length up to some cut-off in time): Given 
the probability $P_t(\theta)\sim \theta^{-\tau}$, $\tau>1$, for the 
walker to reach the wall for the first time at $t_a=\theta$ (at which
point the walk disappears without further contributing to the statistics
in the numerical simulation), we have the identity
\begin{eqnarray}
C(t+t_w,t_w)&=&\int_0^{\infty} d\theta C(t+t_w,t_w|\theta)
P_t(\theta)\cr
\noalign{\medskip}
&\sim& C^{\rm intr}(t+t_w,t_w)\left(t_w+ t\right)^{1-\tau},
\end{eqnarray}
and thus
\begin{eqnarray}
C^{\rm intr}(t+t_w,t_w)&\sim& C(t+t_w,t_w)\left(t_w+ t\right)^{\tau-1}\cr
\noalign{\medskip}
&\sim& t^{-{1\over 2}} f\left({t\over t_w}\right) \left(1+{t\over
t_w}\right)^{\tau-1}.
\end{eqnarray}
Hence, the correct scaling function for the aging behavior of the
intrinsic process is given by 
\begin{eqnarray}
f^{\rm intr}(x)\sim f(x) (1+x)^{\tau-1},
\end{eqnarray}
with $f^{\rm intr}(x\ll 1)\sim 1$ and $f^{\rm intr}(x\gg 1)\sim
x^{\tau-2}$. Of course, $\tau=3/2$ from the familiar first-passage time
of a random walk \cite{Hughes}, and aging remains intact although the 
cross-over in $f^{\rm intr}$ is less dramatic then before for $f$. In Fig.~2 
we plot results for $C^{\rm intr}$ from numerical simulations, and in Fig.~3 
we plot the scaling $f^{\rm intr}$ for the data in Fig.~2.

As mentioned before, the stochastic annihilation process $A+A\to 0$ is
closely related to the random walk model with an absorbing wall and,
indeed, the intrinsic scaling behavior  found here (aside from an
overall factor of $\sqrt{t}$) coincides with the on
reported in Eq.~(9) of Ref.~\cite{FKR} (for $t\to t+t_w$ and $\xi\to
t/(t+T_w)$). 

While in this walk model the aging behavior  remains intact even for the
intrinsic properties of the process, it is important to note that in some cases
the observed aging behavior can be entirely attributed to improper
normalization of the correlation functions in a process in which the
norm depletes. Of course, such a situation can not be considered as
aging behavior. (In fact, in a recent paper \cite{star} this effect has even
been proposed as a general explanation for aging.)

\section{Conclusions and Acknowledgments}

We have shown that a simple, solvable model of a random walk near a wall
can exhibit aging behavior which illuminates many features that lead to
aging behavior in more complicated (disordered) systems, using a
domain-growth picture \cite{F+H}. Of course, systems with many
interacting degrees of freedom such as spin glasses or folding proteins
exhibit a nontrivial phase space structure \cite{slowdyn} which leads to slow
relaxation and aging behavior. Their dynamics is described merely on a
coarse, phenomenological level by such a simple model. But the
connection between the micro-dynamics and the macroscopic phenomena is
not only beyond the scope of this article, but as well itself very much
under development still. Instead of ``explaining'' experimental or
theoretical results, this random walk model is meant to illustrate some
of the questions involved. (After all, it is still rare to consider
situations with broken time-translational invariance, and thus violations of
the fluctuation dissipation theorem, without which two-time correlation
functions would be redundant.) Furthermore, we have discussed some of the 
pitfalls in identifying aging behavior in systems which do not conserve 
the norm, and how to obtain the intrinsic features of such systems.

I would like to thank Maya Paczuski for discussing some of the issues
considered, and Eli Ben-Naim for a critical reading of the manuscript.

\figure{
Log-log plot of the two-time correlation function $C(t+t_w,t_w)$ 
(arbitrary scale) from numerical simulations
of a walk near a reflecting wall. 
Each plot contains data for $2^{i-1}\le t_w<2^i$ for $2\le
i\le 14$ where $i$ labels each graph from bottom to top. Initially, 
for $t<t_w$, each correlation function falls like $1/\sqrt{t}$ with a
crossover at $t\sim t_w$, after which the walk notices the effect of 
the (domain-)wall and the function falls like $2/\sqrt{t}$ for $t\gg
t_w$. The continuous line is inserted to guide the eye and show that
the plot of $C$ for each $t_w$ indeed falls like $t^{-1/2}$ in two separate
regimes which only differ by a factor of $2$.}

\figure{
Log-log plot of the normalized, intrinsic two-time correlation function
$C^{\rm intr}(t+t_w,t_w)$  from numerical simulations
of a walk near an absorbing wall. (Each plot is shifted to avoid
overlaps.) Each plot contains data for $8^{i-1}\le t_w<8^i$ for $1\le
i\le 5$ where $i$ labels each graph from bottom to top. Initially, 
for $t<t_w$, each correlation function falls like $1/\sqrt{t}$ with a
crossover at $t\sim t_w$, after which the effect of the (domain-)wall
becomes noticeable and the function falls like $1/t$.}

\figure{
Scaling plot $f(t/t_w)\sim \sqrt{t}~C^{\rm intr}(t+t_w,t_w)$ as a
function of the scaling variable $t/t_w$ for the data in Fig.~2. All
data collapses reasonably well onto a single scaling graph which is
constant for small argument and falls like an inverse square-root for
large argument (such as the dashed line drawn for reference).}

\begin{references}
\bibitem{rem}
There are only two generic cases regarding the asymptotic behavior
considered here: totally reflecting boundary conditions
on one side, and totally or partially absorbing boundary conditions on
the other.
\bibitem{rem2}
The asymptotic behavior does not depend on the starting position due to
the diffusive nature of the process.
\end{references}
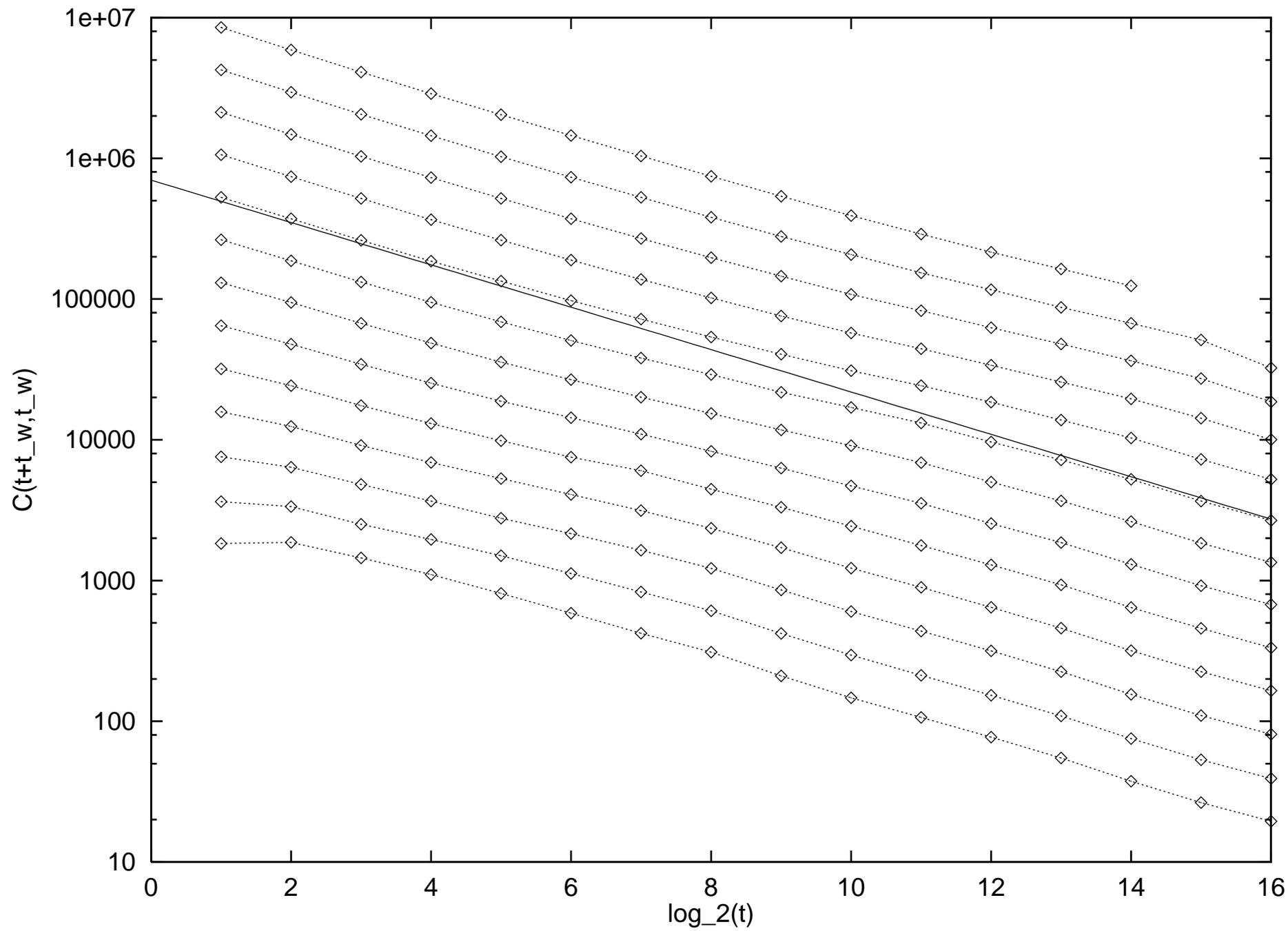
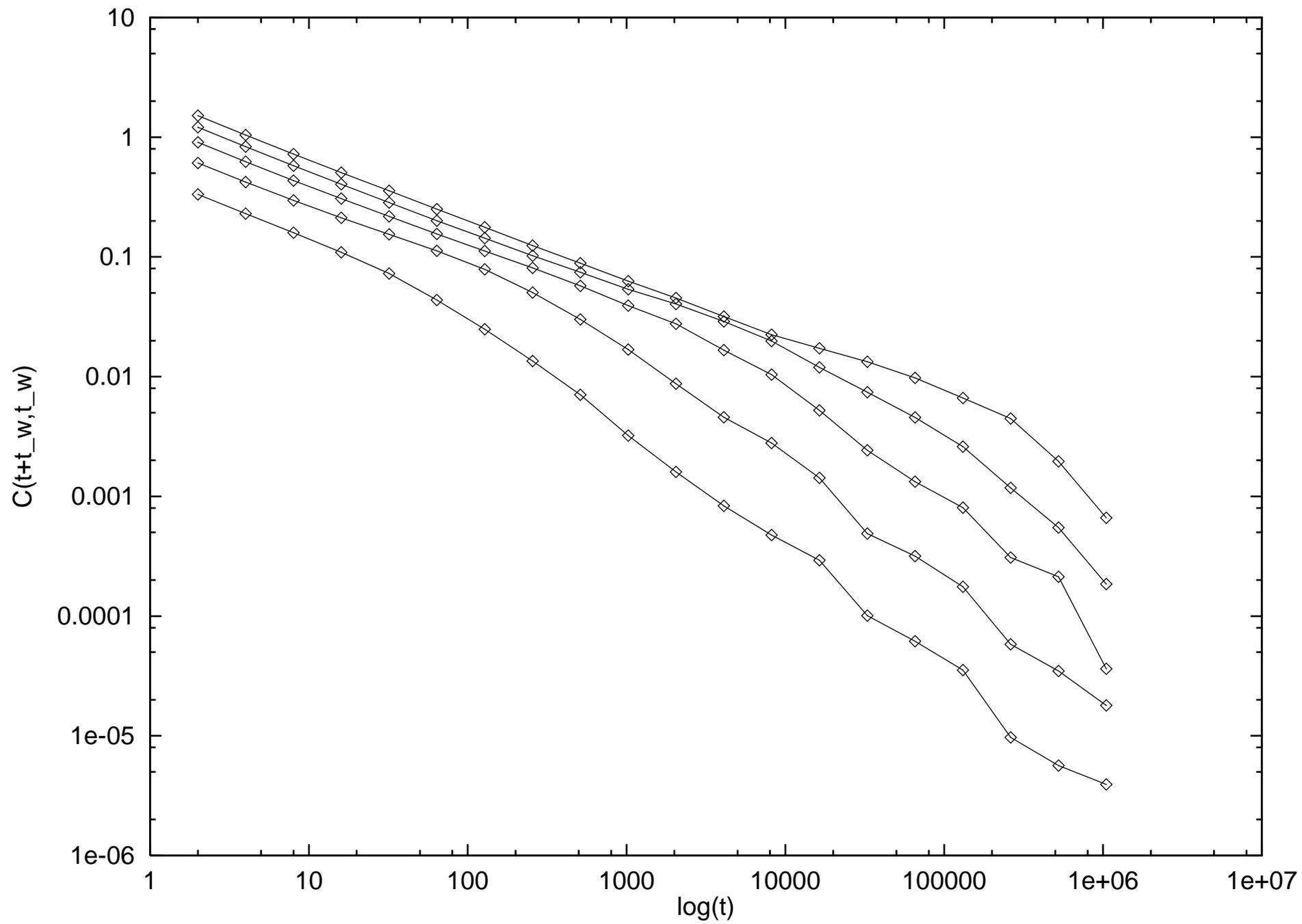
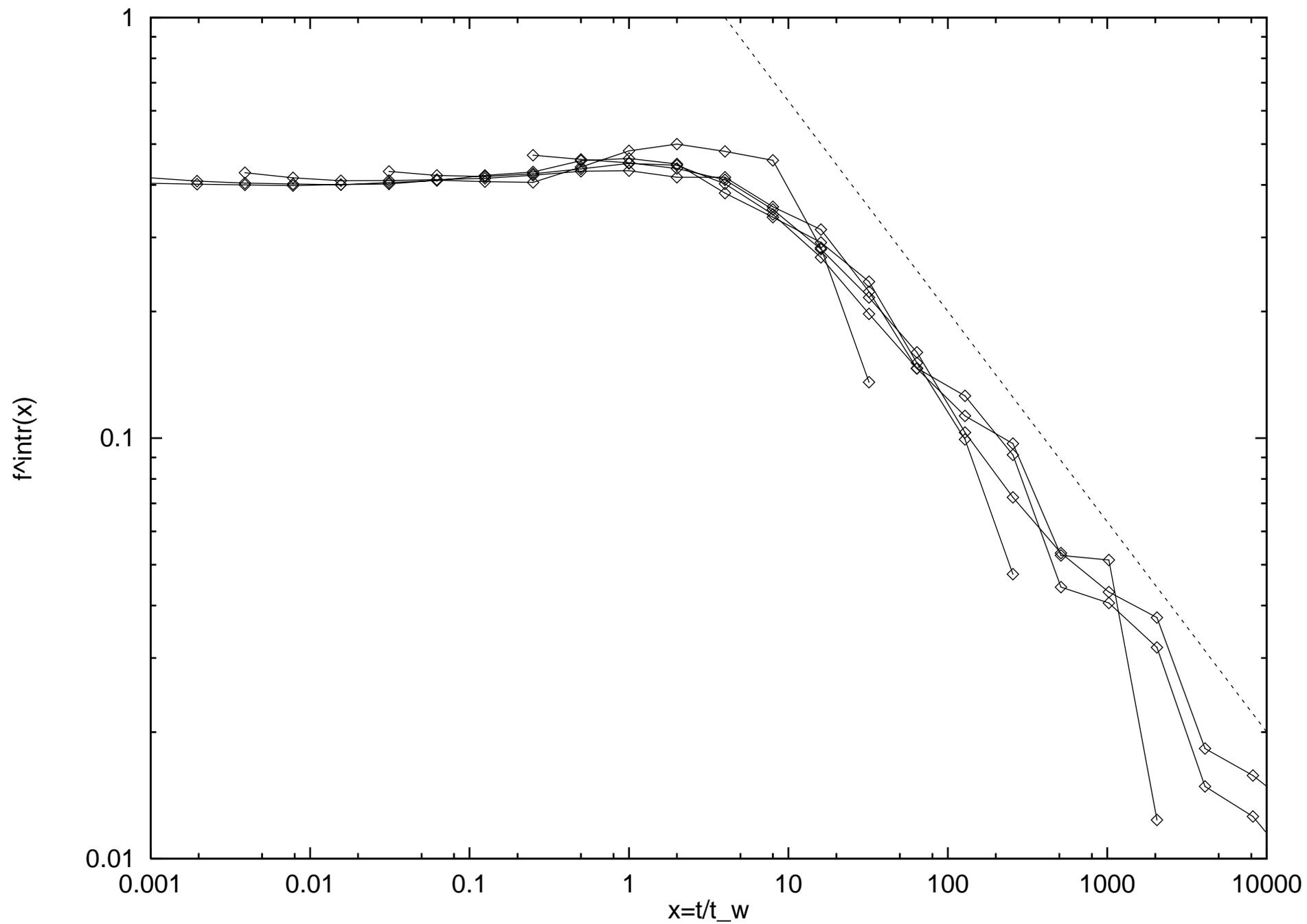
\end{document}